\documentclass[12pt]{article}
\usepackage{amsfonts,amsmath}

   \font\tenmb   =cmmib10 scaled 1400            

\def\boldmath{\tenmb}

\def\nn{\nonumber}
\def\be{\begin{equation}}
\def\eqn#1{\be\label{#1}}
\def\ee{\end{equation}}

\def\bea{\begin{eqnarray}}
\def\eqnn#1{\bea\label{#1}}
\def\eea{\end{eqnarray}}

\def\bbc{{C\kern-7.5pt I}}
\def\bbcc{{C\kern-9.5pt I}}

\def\s{{\sigma}}
\def\ta{{\tilde{\alpha}}}
\def\white#1{\mathop{\bigcirc}\limits_{#1}}

\def\gray#1{\mathop{\bigotimes}\limits_{#1}}
\def\riga{-\kern-4pt - \kern-4pt -}
  \def\cc{{\cal C}}
  
\def\cg{{\cal G}}

\def\k{{\kappa}} \def\d{{\delta}} \def\a{{\alpha}}
\def\o{{\bar 0}} \def\I{{\bar 1}}

\begin{document}

\begin{center}

 {\LARGE {\textbf{ Note on Centrally Extended
$\boldmath{su(2/2)}$ and Serre Relations}}}

\vspace{10mm}

{\bf \large V.K. Dobrev }

\vspace{4mm}

  Institute for Nuclear Research and Nuclear Energy,\\
Bulgarian Academy of Sciences,\\ 72 Tsarigradsko Chaussee, 1784
Sofia, Bulgaria \end{center}

\vspace{6mm}

\begin{abstract}
We point out that the nontrivial central extension of the superalgebra
$su(2/2)$ is related to the some not so well-known Serre relations.
\end{abstract}

\vskip 2cm

\section{Introduction}

Recently, the superalgebra ~$su(2/2)$~ played an important role in
the study of planar N=4 super Yang-Mills and in the construction of
S-matrix for dynamic spin chains, cf. \cite{Beisert}, and for latest
references, \cite{Ines}.  In \cite{Beisert} was  exploited the fact
that the superalgebra ~$su(2/2)$~ has an extraordinary feature,
namely, it is the only basic classical Lie superalgebra\footnote{A
basic classical Lie superalgebra is a Lie superalgebra which has a
non-degenerate, even, supersymmetric, invariant bilinear form
\cite{Kac}.} that has nontrivial central extension \cite{IohKog} (besides
the trivial extension of $psu(n/n)$ to $su(n/n)$). This
feature is related to the fact that the nontrivial central extension of the
complexification $sl(2/2;\bbc)$ may be obtained by contraction from
the algebra $D(2,1;\bar\s)$.

The aim of this Note is to point out that the origin of the above
phenomenon is in some not so well-known Serre relations for
$sl(2/2;\bbc)$. Actually, we pose the question more generally in
order to stress how special the superalgebra $sl(2/2;\bbc)$ is.

Serre relations \cite{Serre} were introduced as an important tool
for handling semi-simple Lie algebras. Using them one can
reconstruct the full algebra starting from the Chevalley generators
which correspond to the simple roots. One would expect that these
relations would be used also for semi-simple Lie superalgebras.
However, that was not the case. Initially,   Lie
superalgebras were given, constructed and classified in the full
Cartan-Weyl basis using their realization in terms of
finite-dimensional super-matrices \cite{Kac}.  Only in the
development of $q$-superalgebras it became necessary to generalize
Serre relations to $q$-deformations of   Lie superalgebras
in order to be able to use Drinfeld's theory of $q$-Serre relations
\cite{Drinfeld}. Initially, it was thought that the generalization
to the $q$-deformed case would be done just by superizing Drinfeld's
$q$-Serre relations.  However, in 1991, several sets of authors
\cite{Kac2,FLV,Scheunert,Yama} discovered independently that for some
superalgebras it is necessary to introduce new Serre relations,
which actually are necessary also in the undeformed case, $q=1$.

The superalgebra $sl(2/2;\bbc)$ is one of the superalgebras that
need  extra Serre relations, yet, it is unique in the sense that
failing to impose them, one obtains a sensible result. Namely, one
obtains the superalgebra $D(2,1;\bar\s)$.

This Note is organized as follows. In the next Section we present
the defining relations for (q-deformed)  superalgebras, including
all Serre relations. Then we specialize to the cases of
$sl(2/2;\bbc)$ and $D(2,1;\bar\s)$ demonstrating their relationship.

\section{Defining relations for (q-deformed)  superalgebras}
\label{required}

  Let  $\cg$ be a complex Lie superalgebra with a symmetrizable
Cartan matrix $A = (a_{jk})$ $= A^d \hat A^s$, where
 $\hat A^s = (\hat a^s_{jk})$ is a symmetric matrix, and $A^d =$~
diag $(\hat d_1,...,\hat d_\ell)$, $\hat d_k\neq 0$.   Then the $q$
- deformation $U_q(\cg)$ of the universal enveloping algebras
$U(\cg)$ is defined \cite{KT} as the associative algebra over $\bbc$
with generators $X^\pm_j ~, ~H_j$~, $j\in J = \{1,\dots ,\ell\}$~
and with relations similar to the even case~: \eqnn{suco}  &[ H_i~,
~ H_j] ~=~ 0 , ~~~[ H_i~, ~X^\pm_j] ~=~ \pm \hat a^s_{ij}~ X^\pm_j
\cr &[ X^+_i~, ~ X^-_j] ~=~ \d_{ij}~ [H_i]_{q_i} ~, \qquad q_i =
q^{\hat d_i} \eea ([,] being the supercommutator), \eqn{suser} ({\rm
Ad}_{q^\k} X^\pm_j)^{n_{jk}}(X^\pm_k) ~=~ 0 ~, ~{\rm for}~ j\neq k
~, ~~\k ~=~ \pm ~; \ee and for every three simple roots, say, $\a_j
~, ~\a_{j \pm 1}$, such that $(\a_j , \a_j) = 0$, $(\a_{j \pm 1} ,
\a_{j \pm 1}) \neq 0$,
 $(\a_{j + 1} , \a_{j - 1}) = 0$, $(\a_j , \a_{j + 1} +
\a_{j - 1}) = 0$, also holds: \eqn{sunu} [ [ X^\pm_j ~,
~X^\pm_{j-1}]_{q^\k} ~,
   ~[ X^\pm_j ~, ~X^\pm_{j+1}]_{q^\k} ] ~=~ 0 \ee
where~: \eqnn{} n_{jk} ~=&~ 1 & {\rm if}\quad \hat a^s_{jj} = \hat
a^s_{jk} = 0 \cr & 2 & {\rm if}\quad \hat a^s_{jj} = 0, ~\hat
a^s_{jk} \neq 0 \cr & 1 - 2\hat a^s_{jk}/\hat a^s_{jj} & {\rm
if}\quad \hat a^s_{jj} \neq 0 \eea  where in
\eqref{suco},\eqref{suser} one uses the deformed supercommutator:
\eqnn{supc} ({\rm Ad}_{q^\k}~X^\pm_j)~(X^\pm_k) ~&=&~ [X^\pm_j ,
X^\pm_k]_{q^\k} ~\equiv~ \cr &&\equiv ~X^\pm_j X^\pm_k ~-~ (-1)^{
p(X^\pm_j)\, p(X^\pm_k) } ~q^{\k (\a_j,\a_k)/2} ~X^\pm_k X^\pm_j
\eea

The above is applicable to the even case, then relations
\eqref{suco},\eqref{suser} for $\k=1$ are the same as for $\k= -1$
and coincide with the usual even Serre relations. The necessity of
the extra relations \eqref{sunu} (including the classical $q=1$
case) was communicated to the author in May 1991 independently by
Scheunert, Kac, and Leites.   In the $q$-deformed case these
relations were written   for $U_q(sl(M/N))$ in \cite{Scheunert} and
 \cite{FLV},  for $U_q(osp(M/2N))$ in \cite{FLV}, and in
general in \cite{Yama}. Here they are given as in \cite{KT}, except
that in \cite{KT} also the external supercommutator is given as
deformed, while in fact it is not. The reason is that the would be
deformation is given by the factor ~$q^{\kappa(\a_j + \a_{j-1},\a_j
+ \a_{j+1})}$, and as can be easily checked the scalar product in
the q-exponent is actually zero. (Later  extra Serre relations
were written also for the affine ($q$-deformed) Kac-Moody
superalgebras \cite{KTb,Yamb} and for any Lie superalgebra
with Cartan matrix \cite{GrLe}. See also recent papers on the
$q$-deformed affine superalgebra  $D(2,1;\bar\s)$ \cite{HSTY} and
the $su(2/2)$ Yangian \cite{ST} relevant in our current context and
related to \cite{Beisert}.)

Pictorially, one case of the situation with the three roots in
\eqref{sunu} is given by the following (part of) Dynkin
diagram: \eqn{sl22} \white{{1}} \riga \gray{{1}} \riga \white{{1}}
\ee where a circle represents an even root, while a crossed (gray)
circle represents an odd root. The number at a node gives the
coefficient with which the corresponding simple root enters the
decomposition of the highest root.

\section{The cases of $\ \boldmath{sl(2/2;\bbcc)}\ $ and
  $\ \boldmath{D(2,1;\bar\s)}\ $}

The three roots  in \eqref{sunu} and three nodes in \eqref{sl22} by
themselves determine the root system and Dynkin diagram of the
superalgebra ~$sl(2/2;\bbc)$. We recall that this superalgebra is
15-dimensional (over $\bbc$), the even subalgebra being
seven-dimensional:
$$ sl(2/2;\bbc)_\o  ~\cong~ sl(2,\bbc) \oplus sl(2,\bbc) \oplus \bbc $$
while the odd part $sl(2/2;\bbc)_\I$ is eight-dimensional. We choose
a distinguished root system (with one odd simple root) corresponding
to \eqref{sl22}, the simple roots being denoted as
~$\a_1,\a_2,\a_3\,$, of which ~$\a_2$~ is odd, the other - even. The
positive roots are: \eqnn{drot} \Delta^+ = \Delta^+_\o \cup
\Delta^+_\I\ , \qquad &\Delta^+_\o = \{ \a_1,\a_3\} \ , \nn\\ &
\Delta^+_\I = \{ \a_2,\a_1+\a_2, \a_2+\a_3, \a_1+\a_2+\a_3\} \eea
(Note that ~$\a_1+\a_2+\a_3$~ is the highest root.)

We denote the Chevalley generators as ~$X^\pm_j\,, H_j\,$,
$j=1,2,3$. Then the Cartan subalgebra of $sl(2/2;\bbc)$ is spanned
by ~$H_j\,$, $j=1,2,3$, though often instead of ~$H_2\,$, (related
to the odd root $\a_2$), is used the central generator ~$K = H_1 +
2H_2 -H_3\,$. (In a matrix realization of $4\times4$ matrices, $K
\sim I_4\,$, the unit $4\times4$ matrix.)

The defining relations \eqref{suco} and \eqref{suser} (for $q=1$)
are clear. We are interested in the new relations \eqref{sunu} which
we write out explicitly for ~$q=1$~: \eqn{sunue} [ [ X^\pm_2 ~,
~X^\pm_{1}]\,,\,   [ X^\pm_2 ~, ~X^\pm_{3}]  ]_+ ~=~ 0 \ .\ee

We pass now to the superalgebra ~$D(2,1;\bar\s)$.  We recall that
this superalgebra is 17-dimensional, the even subalgebra being
nine-dimensional:
$$ sl(2/2;\bbc)_\o  ~=~ sl(2,\bbc) \oplus sl(2,\bbc) \oplus sl(2,\bbc)$$
while the odd part $sl(2/2;\bbc)_\I$ is eight-dimensional
\cite{Kac}. We choose a distinguished root system, the simple roots
being denoted as ~$\a_1,\a_2,\a_3\,$, of which ~$\a_2$~ is odd, the
other - even. The positive roots are: \eqnn{droot} \Delta^+ =
\Delta^+_\o \cup \Delta^+_\I\ , \qquad &\Delta^+_\o = \{
\a_1,\a_3,\a_1+2\a_2+\a_3\} \ , \nn\\  &\Delta^+_\I = \{
\a_2,\a_1+\a_2, \a_2+\a_3, \a_1+\a_2+\a_3\} \eea We note that ~$\ta
~=~ \a_1+2\a_2+\a_3$~ is the highest root, and the root system
corresponds to the following Dynkin diagram: \eqn{dynd21}
\white{{1}} \longleftarrow \gray{{2}}\longrightarrow \white{{1}} \ee

For this superalgebra the defining relations are only \eqref{suco}
and \eqref{suser} for $q=1$, since the Dynkin diagram is not of the
type \eqref{sl22}. Our interest is to give an explicit expression
for the Cartan-Weyl generators ~$X^\pm_\ta$~ corresponding to the
highest root ~$\tilde{\a}\,$. After a simple calculation we obtain:
\eqn{highr} X^\pm_\ta ~=~ [ [ X^\pm_2 ~, ~X^\pm_{1}]\,,\,   [
X^\pm_2 ~, ~X^\pm_{3}]  ]_+ \ee

Identifying the simple roots in \eqref{drot} and \eqref{droot} and
comparing \eqref{sunue} and \eqref{highr} we see that the
superalgebras ~$sl(2/2;\bbc)$, ~$D(2,1;\bar\s)$  differ by the extra
Serre relations \eqref{sunue} which are needed for ~$sl(2/2;\bbc)$~
but are absent for ~$D(2,1;\bar\s)$.  Consequently, we can pass from
~$D(2,1;\bar\s)$~ to ~$sl(2/2;\bbc)$~ by a contracting procedure in
which the root vectors ~$X^\pm_\ta$~ \eqref{highr} are sent to zero.
However, we may use a more general contraction procedure, as in
\cite{Beisert}, in which the root vectors are replaced by central
elements, (not necessarily zero), i.e., setting: \eqn{sunuee}
X^\pm_\ta ~\longrightarrow~ \cc^\pm \in \bbc \ee we obtain the
central extension of ~$sl(2/2;\bbc)$~ found in \cite{IohKog}.

\section*{Acknowledgements}
The author would like to thank for hospitality the Service de
Physique Th\'eorique, C.E.A.-Saclay, and for discussions and
information on the literature  V. Kazakov, I.
Kostov, D. Leites, D. Serban, F. Spill, A. Torrielli, and H. Yamane.
 The author was  supported in part by
  the European RTN network {\em ``Forces-Universe''} (contract
No.\textsl{MRTN-CT-2004-005104}), by Bulgarian NSF grant  \textsl{DO
02-257}, and by the Alexander von Humboldt Foundation in the
framework of the Clausthal-Leipzig-Sofia Cooperation.

\end{document}